\documentclass[showpacs,preprintnumbers,amsmath,amssymb]{revtex4} % insert this command in the brackets: twocolumn,
\usepackage{graphicx}% Include figure files
\usepackage{dcolumn}% Align table columns on decimal point
\usepackage{bm}% bold math
\usepackage{natbib}
\usepackage{amssymb,amsmath}
\usepackage{longtable}
\usepackage{txfonts}
\usepackage{romannum}

\bibpunct{(}{)}{;}{a}{}{,}

\usepackage[a4paper=true,dvipdfm=true,pagebackref=true]{hyperref}
\hypersetup{pdftitle = The title of my PDF, pdfauthor = My name, pdfsubject= The subject, pdfkeywords = keyword1 keyword2 keyword3}
\hypersetup{colorlinks = true, linkcolor = green, anchorcolor = red, citecolor = blue, filecolor = red, pagecolor = red, urlcolor = red}
\newcommand{\aap}{Astronom. and Astrophys.}
\newcommand{\pasa}{Publications of the Astronomical Society of Australia}
\newcommand{\aaps}{Astronom. and Astrophys. Suppl. Ser.}
\newcommand{\aj}{Astronom. J.}
\newcommand{\apjs}{Astrophys. J. Suppl.}
\newcommand{\mnras}{Monthly Notices Roy. Astronom. Soc.}
\newcommand{\pasp}{Publ. Astronom. Soc. Pacific}
\begin{document}

\title{Physical and Geometrical Parameters of CVBS \Romannum{12}: Fin 350 (Hip 64838)}
\author{\firstname{M.\,A.}~\surname{Al-Wardat}}
\email{mwardat@aabu.edu.jo}
\affiliation{Department of Physics, Al al-Bayt University, PO Box: 130040, Mafraq, 25113, Jordan.}
\author{\firstname{J.\, A.}~\surname{Docobo}}
%\email{}
\affiliation{Observatorio Astron\'{o}mico Ram\'{o}n Mar\'{i}a Aller, Universidade de Santiago de Compostela, Avenida das Ciencias s/n, 15782 Santiago de Compostela, Spain}
\author{\firstname{A.\, A.}~\surname{ Abushattal}}
\affiliation{Observatorio Astron\'{o}mico Ram\'{o}n Mar\'{i}a Aller, Universidade de Santiago de Compostela, Avenida das Ciencias s/n, 15782 Santiago de Compostela, Spain}
\author{\firstname{P.\,  P.}~\surname{ Campo}}
\affiliation{Observatorio Astron\'{o}mico Ram\'{o}n Mar\'{i}a Aller, Universidade de Santiago de Compostela, Avenida das Ciencias s/n, 15782 Santiago de Compostela, Spain}

\received{October 24, 2016}%
\revised{January 10, 2017}%

%\date{\today}

\begin{abstract}
A complete astrophysical and dynamical study of the close visual binary system (CVBS)  (A7V + F0V), Finsen 350, is presented.
Beginning with the entire observational spectral energy distribution (SED) and the magnitude difference between the subcomponents, Al-Wardat's complex method for analyzing close visual binary stars (CVBS) was applied as a reverse method of building the individual and entire synthetic SEDs of the system. This was combined with Docobo's analytic method to calculate the new orbits.
Although possible short ($\approx$ 9 years) and long period ($\approx$ 18 years) orbits could be considered taking into account the similar results of the stellar masses obtained for each of them (3.07 and 3.41 $M_{\odot}$, respectively), we confirmed that the short solution is correct.
In addition, other physical, geometrical and dynamical parameters of this system such as the effective temperatures, surface gravity accelerations, absolute magnitudes, radii, the dynamical parallax, etc., are reported. The main sequence phase of both components with age around 0.79 Gy is approved.

\end{abstract}

%\keywords{stars: binaries: visual; stars: interferometric binaries; stars: photometry}

\maketitle
\section{Introduction}

Most of what look like single stars in the sky are, actually,  binary or multiple systems, as revealed by Hipparcos mission \citep{2002A&A...385...87B,1998AstL...24..673S}.
In general,  stellar binary systems represent the key source of  stellar parameters especially masses and distances. As for subgiant stars and the lower part of the main sequence, they practically define our understanding of stellar physical properties \citep{2001A&A...366..868D}.

Except for eclipsing binary stars, there is no direct way to measure the physical and geometrical parameters of stellar binary systems, even with the aid of modern techniques of observation, such as speckle interferometry and adaptive optics. The situation is a bit complicated in the case of the close visual binary stars (CVBS), especially subgiants, which represent a small stellar category due to their short evolutionary phase \citep{2006essp.book.....S}.

%Due to the expansion, the outer layers cool and envelope opacity increases drastically. As a consequence energy trapping in the envelope becomes increasingly efficient, and supports the expansion of the outer layers. During this phase the structure moves in the HRD from the blue to the red side at almost constant surface luminosity. This is called Sub-Giant Branch (SGB). The evolutionary rate during this phase is roughly the Kelvin¨CHelmholtz timescale, being of the order of ¡«12 Myr for a 3M and ¡«1 Myr for a 6M. This evolutionary phase is so short for stars of high and intermediate mass that the chance of observing objects in this phase is very small. This leads to the presence of the so called Hertzsprung gap (a lack of stars along the SGB) in the HRD of stellar systems populated by intermediate-mass and massive stars during the post-MS phases. \citep{2006essp.book.....S}

This difficulty was overcome by applying Al-Wardat's complex method for analyzing CVBS, which combines different observational results and analytical techniques such as speckle interferometry, spectrophotometry, atmospheric modelling, and dynamical analysis. This method yields an accurate determination of the complete set of physical and geometrical parameters which include effective temperatures, gravity accelerations, radii, masses, orbital parameters, absolute magnitudes, densities, spectral types, and luminosity classes of the components of the CVBS. First devised by \cite{2002BSAO...53...51A, 2007AN....328...63A}, the method was applied to several main sequence CVBS such as ADS 11061, COU 1289, COU 1291, HIP 11352, HIP 11253, HIP 70973, and HIP 72479 \citep{2002BSAO...53...51A, 2007AN....328...63A, 2009AN....330..385A, 2012PASA...29..523A, 2009AstBu..64..365A} as well as to the subgiant CVBS, HD 25811, HD 375 and HD 6009 (\cite{2014AstBu..69...58A}, \cite{2014PASA...31....5A}, \cite{2014AstBu..69..454A}). It was also applied to the spectroscopic CVBS Gliese 762.1 (Paper X in this series).

As a consequence of the previous work, this paper (the \Romannum{12}  in its series)  presents the  analysis of the CVBS FIN 350 (HIP64838). It was  reported as a double star in the Bright Star Catalogue and by \cite{1963icvd.book.....J} in the Index Catalogue of Visual Double Stars. This star was observed for the first time in 1959.47 by Finsen who established it as a binary and measured a separation ($\rho$) of $0.1$ arcsec, with a position angle ($\theta$) of $27^{\circ}$. He also determined that both components were of the same brightness, 7.1 mag, where he used an eyepiece interferometer developed by himself at that time for his observations.

It is not clear yet wether the components of the  system belong to the main sequence or to the  subgiant phase.   \cite{1975AJ.....80..637M} assigned the system to the F0V MK spectral type,  while \cite{1976PASP...88...95C} assigned it to A7IV (or mild A7m). So, it seems an interesting system, and the estimated parameters in this work will enhance our knowledge of binary systems in general, and will consequently help to understand the formation and evolution mechanisms of such systems.

% (mentioned in the results)The spectral type and luminosity class of the system were designated as F0V by \cite{1975AJ.....80..637M} and  A7IV (or mild A7m:) by \cite{1976PASP...88...95C}. % A6V by \cite{1995ApJS...99..135A} with $v sin(i)=120\textrm{km}s^{-1}$.

\begin{table}
\begin{center}
\caption{Data from SIMBAD and NASA/IPAC}
 \label{stardata}
\begin{tabular}{@{}l|c|c}
\hline \hline
 \  & FIN 350 & Source\\
 \hline
$\alpha_{2000}$ & $13^h17^m29^s.853$     & SIMBAD\\
$\delta_{2000}$  &   $ -00^h40^m33^s.83$ & SIMBAD      \\
 HD           & 115488& SIMBAD       \\
 Tych  & 4958-1448-1 & SIMBAD \\
 $m_v$  & $6^m.358$ & SIMBAD \\
  E(B-V)         & 0.026     & NASA/IPAC \\
Av & 0.128       & \cite{2005oasp.book.....G}  \\
$V_J$(Hip) & $6^m.36\pm0.06$ & HIP97 \\
$(B-V)_J$(Hip) & $0^m.26\pm0.06$ & HIP97 \\
$B_T$        &  $6^m.66\pm0.004$ & TYCHO \\
$V_T$  &  $6^m.380\pm0.004$& TYCHO \\
  $(B-V)_J$(Tych)  & $0^m.255\pm0.05$ & TYCHO \\
 $\pi_{Hip} $(new)   & $12.28\pm0.77$\ mas  & HIP07 \\
 $\pi_{Hip} $(old)  & $13.45\pm0.77$\ mas & HIP97 \\
   \hline \hline
\end{tabular}\\
NASA/IPAC: http://irsa.ipac.caltech.edu\\
SIMBAD: http://simbad.u-strasbg.fr\\
HIP97: \cite{HIPshort}\\
HIP07: \cite{2007A&A...474..653V}\\
TYCHO: \cite{2000A&A...355L..27H}
\end{center}
\end{table}

Table~\ref{stardata} contains basic data of this system from SIMBAD, NASA/IPAC, Hipparcos and Tycho Catalogues (ESA).

\section{Orbital Elements}

Previous orbits for this system were calculated by \cite{1988A&AS...74..507B}: P=8.989 yr, a=86 mas.; \cite{1996AJ....111..370H}: P=9.046 yr, a=79.7 mas.; and more recently by \cite{2011AJ....141..180H}: P=9.156 yr, a=80.8 mas.

The latest speckle observations suggested a revision of the earlier solutions and, taking into account the small difference of magnitude between the components, we decided to calculate not only the short-period orbit ($\approx$ 9 yr) but also the long-period orbit ($\approx$ 18 yr) twisting several measurements by 180$^{\circ}$ (see Figures 1 and 2). Concretely, the orbit of Horch et al. provides the following residuals in $\theta$ for the measurements of 2007.329, 2009.260, and 2009.446 (3 observations in this last epoch), which are near the periastron passage: +4$^{\circ}$7, +3$^{\circ}$0, +4$^{\circ}$0, +5$^{\circ}$8, and +5$^{\circ}$4. Moreover, Horch himself provided us with the unpublished observations performed in 2013.4026. Docobo's analytic method \citep{1985CeMec..36..143D, 2012ocpd.conf..119D} was used to calculate the new orbits.

\begin{center}
\tiny
\begin{longtable}{lcccccccccccc}
%Here is the caption, the stuff in [] is the table of contents entry,
%the stuff in {} is the title that will appear on the first page of the
%table.
\caption[Observational data of Finsen 350]{Observational data of Finsen 350.} \label{deltam2} \\

%This is the header for the first page of the table...
\hline \hline \\[-2ex]
   \multicolumn{1}{c}{Date} &   \multicolumn{1}{c}{$\theta(deg)$} &   \multicolumn{1}{c}{$\sigma\theta(deg)$}
   &   \multicolumn{1}{c}{$\rho(arcsec)$}&   \multicolumn{1}{c}{$\sigma\rho(arcsec)$}&   \multicolumn{1}{c}{$\delta m$}
   &   \multicolumn{1}{c}{$\sigma\delta m$}&   \multicolumn{1}{c}{Filter $\lambda$}&   \multicolumn{1}{c}{$\delta\lambda$}&   \multicolumn{1}{c}{Tel.}&   \multicolumn{1}{c}{Ref.*}&   \multicolumn{1}{c}{Meth}&   \multicolumn{1}{c}{weight}
    \\[0.5ex] \hline
   \\[-1.8ex]
\endfirsthead

%This is the header for the remaining page(s) of the table...
\multicolumn{13}{c}{{\tablename} \thetable{} -- Continued} \\[0.5ex]
  \hline \hline \\[-2ex]
  \multicolumn{1}{c}{Date} &   \multicolumn{1}{c}{$\theta(deg)$} &   \multicolumn{1}{c}{$\sigma\theta(deg)$}
   &   \multicolumn{1}{c}{$\rho(arcsec)$}&   \multicolumn{1}{c}{$\sigma\rho(arcsec)$}&   \multicolumn{1}{c}{$\delta m$}
   &   \multicolumn{1}{c}{$\sigma\delta m$}&   \multicolumn{1}{c}{Filter $\lambda$}&   \multicolumn{1}{c}{$\delta\lambda$}&   \multicolumn{1}{c}{Tel.}&   \multicolumn{1}{c}{Ref.*}&   \multicolumn{1}{c}{Meth}&   \multicolumn{1}{c}{weight}
     \\[0.5ex]\hline
\endhead

%This is the footer for all pages except the last page of the table...
  \multicolumn{13}{l}{{Continued on Next Page\ldots}} \\
\endfoot

%This is the footer for the last page of the table...
  \\[-1.8ex] \hline \hline
\endlastfoot

%Now the data..
  1959.47&	     27.1	&	-	&	0.131	&	-	&	0.0	&	-	&	-	 &	 -	&	 0.7 	 &	 \cite{1959MNSSA..18...31F}  	 &	J& 5	\\
  1960.55&  	29.2	&	-	&	0.126	&	-	&	0.1	&	-	&	-	 &	 -	&	 0.7  	 &	 \cite{1961CiUO..120..367F}	 &	 J  & 5	\\
  1964.530& 	162.5	&	-	&	0.104	&	-	&	0.0	&	-	&	-	 &	 -	&	 0.7  	 &	 \cite{1965ROCi..124...79F} 	 &	J & 5	\\
  1965.545& 	186.3	&	-	&	0.126	&	-	&	0.0	&	-	&	-	 &	 -	&	 0.7  	 &	 \cite{1965ROCi..124...79F}  	 &	J & 5	\\
  1966.520& 	201.5	&	-	&	-	    &	-	&	-	&	-	&	-	 &   -   &	 0.7	 &   \cite{1967ROCi..126..138F} 	 &	J & 0	\\
  1966.527& 	203.2	&	-	&	0.134	&	-	&	0.0	&	-	&	-	 &	 -	&	 0.7  	 &	 \cite{1967ROCi..126..138F} 	 &	J & 5 	\\
  1967.5447&    208.4	&	-	&	0.140	&	-	&	0.0	&	-	&	-	 &	 -	&	 0.7  	 &	 \cite{1969ROCi..128..187F}  	 &	J & 5	\\
  1968.545&	    200.8	&	-	&	0.137	&	-	&	0.0	&	-	&	-	 &	 -	&	 0.7  	 &	 \cite{1969ROCi..128..187F} 	 &	J & 5	\\
  1976.2959&	12.5	&	0.7	&	0.131	&	0.001&	-	&	-	 &	552	 &	 20	&	 3.8   	 &	 \cite{1978ApJ...225..932M} 	 &	Sc &	15\\
  1976.3697&	13.9	&	-	&	0.114	&	-	&	-	&	-	&	552	 &	 20	&	 2.1   	 &	 \cite{1982ApJS...49..267M} 	 &	Sc &	10\\
  1976.4570&	15.0	&	0.4	&	0.129	&	0.001&	-	&	-	 &	552	 &	 20	&	 3.8   	 &	 \cite{1978ApJ...225..932M}  	 &	Sc & 15	\\
  1977.0877&	18.6	&	0.6	&	0.131	&	0.001&	-	&	-	 &	552	 &	 20	&	 3.8   	 &	  \cite{1979ApJ...228..493M}	 &	Sc & 15	\\
  1977.1751&	15.9	&	-	&	0.120	&	-	&	-	&	-	&	552	 &	 20	&	 2.1   	 &	 \cite{1982ApJS...48..273M} 	 &	Sc	& 10\\
  1977.3280&	19.9	&	-	&	0.119	&	-	&	-	&	-	&	552	 &	 20	&	 2.1   	 &	 \cite{1982ApJS...48..273M} 	 &	Sc & 10	\\
  1978.1499&	26.9	&	0.5	&	0.118	&	0.001&	-	&	-	 &	470	 &	 -	&	 3.8	 &	  \cite{1980ApJS...43..327M} 	 &	Sc	& 15\\
  1978.3109&	32.3	&	-	&	0.112	&	-	&	-	&	-	&	470	 &	 -	&	 2.1	 &	  \cite{1984cimb.book.....M} 	 &	Sc & 10	\\
  1979.3622&	38.3	&	-	&	0.085	&	-	&	-	&	-	&	470	 &	 -	&	 3.8	 &	  \cite{1982ApJS...49..267M}	 &	Sc	& 15\\
  1983.0701&	350.3	&	-	&	0.080	&	-	&	-	&	-	&	549	 &	 22	&	 3.8   	 &	  \cite{1987AJ.....93..688M} 	 &	Sc &	15\\
  1983.4332&	355.4	&	-	&	0.093	&	-	&	-	&	-	&	549	 &	 22	&	 3.8   	 &	 \cite{1987AJ.....93..688M} 	 &	Sc &	15\\
  1984.0532&	1.1	   &	-	&	0.125	&	-	&	-	&	-	&	549	 &	 22	&	 3.8   	 &	 \cite{1987AJ.....93..688M} 	 &	Sc &	15\\
  1984.3752&	5.6	   &	-	&	0.115	&	-	&	-	&	-	&	549	 &	 22	&	 3.8   	 &	 \cite{1987AJ.....93..688M} 	 &	Sc	& 15\\
  1984.3807&	5.3	   &	-	&	0.116	&	-	&	-	&	-	&	549	 &	 22	&	 3.8   	 &	 \cite{1987AJ.....93..688M} 	 &	Sc & 15	\\
  1984.3835&	5.9	   &	-	&	0.116	&	-	&	-	&	-	&	549	 &	 22	&	 3.8   	 &	 \cite{1987AJ.....93..688M} 	 &	Sc & 15	\\
  1985.1805&	13.0	&	-	&	0.117	&	-	&	-	&	-	&	600	 &	 14	&	 6.0   	 &	 \cite{1987PAZh...13..508B}	 &	 S  &	20\\
  1985.2438&	14.0	&	2.9	&	0.126	&	0.013&	-	&	-	 &	625	 &	 75  	&	 1.9	&	\cite{bonneau1986speckle}	 &	 S  & 10\\
  1985.3389&	10.5	&	-	&	0.129	&	-	&	-	&	-	&	549	 &	 22	&	 3.0   	 &	 \cite{2000AJ....119.3084H} 	 &	Sc & 15	\\
  1985.4840&	13.8	&	-	&	0.126	&	-	&	-	&	-	&	549	 &	 22	&	 3.8   	 &	\cite{1987AJ.....93..688M} 	 &	Sc	& 15\\
  1986.4067&	20.3	&	-	&	0.127	&	-	&	-	&	-	&	549	 &	 22	&	 3.8   	 &	\cite{1989AJ.....97..510M}	 &	 Sc 	& 15\\
  1987.2642&	26.2	&	-	&	0.117	&	-	&	-	&	-	&	549	 &	 22	&	 3.8   	 &	\cite{1989AJ.....97..510M}	 &	 Sc 	& 15\\
  1987.3800&	27.5	&	-	&	0.108	&	-	&	-	&	-	&	-	 &	 -	&	 6.0	 &	\cite{1989AISAO..28..107B} 	 &	 S  & 20	\\
  1987.3800&	27.5	&	-	&	0.108	&	-	&	-	&	-	&	-	 &	 -	&	 6.0	 &	\cite{1991AISAO..31...80B}  	 &	 S	& 20 \\
  1988.1655&	35.5	&	-	&	0.088	&	-	&	-	&	-	&	549	 &	 22	&	 3.6   	 &	\cite{1993AJ....106.1639M} &	 Sc & 15 \\
  1988.2524&	36.8	&	-	&	0.080	&	-	&	-	&	-	&	549	 &	 22	&	 3.8   	 &	\cite{1989AJ.....97..510M}	 &	 Sc & 15	\\
  1990.2759&	33.8	&	-	&	0.053	&	-	&	-	&	-	&	467	 &	 16	&	 3.8   	 &	 \cite{1992AJ....104..810H}  	 &	Sc	& 15 \\
  1991.3186&	331.0	&	-	&	0.059	&	-	&	-	&	-	&	549	 &	 22	&	 3.8   	 &	\cite{1994AJ....108.2299H}	 &	 Sc 	&  15 \\
  1992.3098&	352.8	&	-	&	0.085	&	-	&	-	&	-	&	549	 &	 22	&	 3.8   	 &	\cite{1994AJ....108.2299H} &	 Sc 	&  15 \\
  1992.4572&	353.9	&	-	&	0.093	&	-	&	-	&	-	&	549	 &	 22	&	 4.0   	 &	 \cite{1996AJ....111..370H} 	 &	Sc &	15\\
  1993.0905&	1.5	   &	-	&	0.105	&	-	&	-	&	-	&	549	 &	 22	&	 4.0   	 &	\cite{1996AJ....111..370H} 	 &	Sc & 15	\\
  1993.1973&	1.6	   &	-	&	0.114	&	-	&	-	&	-	&	549	 &	 22	&	 3.8   	 & \cite{1994AJ....108.2299H} &	 Sc & 15	\\
  1995.1495&	17.8	&	-	&	0.120	&	-	&	-	&	-	&	549	 &	 22	&	 2.5   	 &  \cite{1997AJ....114.1639H}	 &	 Sc & 10	\\
  1995.3109&	17.2	&	-	&	0.121	&	-	&	-	&	-	&	549	 &	 22	&	 2.5   	 &	 \cite{1997AJ....114.1639H}	 &	 Sc & 10	\\
  1996.1840&	24.2	&	-	&	0.116	&	-	&	-	&	-	&	549	 &	 22	&	 4.0   	 &	 \cite{2000AJ....119.3084H} 	 &	Sc	& 15\\
  2001.2708&	350.3	&	0.8	&	0.078	&	0.003&	-	&  -  &	550	&	 14	 &	6.0   	 &	\cite{2006BSAO...59...20B} 	 &	S & 20	\\
  2001.2708&	350.2	&	0.6	&	0.079	&	0.003&	-	&	-	 &	600	 &	 30	&	 6.0   	 &	 \cite{2006BSAO...59...20B} 	 &	S & 20	\\
  2001.2708&	350.3	&	0.7	&	0.078	&	0.003&	-	&	-	 &	750	 &	 35	&	 6.0   	 &	 \cite{2006BSAO...59...20B} 	 &	S & 20	\\
  2002.3224&	1.8	    &	-	&	0.100	&	-	&	0.62	&	-	&	 550	&	 40	 &	3.5	&	 \cite{2008AJ....135.1334H}	 &	S & 15	\\
  2002.3224&	3.1	   &	-	&	0.106	&	-	&	0.48	&	-	&	 754	&	 44	 &	3.5	&	 \cite{2008AJ....135.1334H}	 &	S & 15	\\
  2004.1960&	11.8	&	0.7	&	0.141	&	0.002&	-	&	-	 &	550	 &	 24	&	 1.55  	&	\cite{2008AJ....135.1334H}	 &	 Su & 10	\\
  2006.1915&	32.1	&	-	&	0.092	&	-	&	-	&	-	&	 550	 &	24	&	 4.0   	&	 \cite{2009AJ....137.3358M}	 &	Su & 15	\\
  2007.0105&	222.2	&	-	&	0.0716	&	-	&	0.99	&	-	 &	 550	&	 40	 &	3.5	&	 \cite{2011AJ....141..180H} 	 &	S & 15	\\
  2007.3286&	238.8	&	-	&	0.0456	&	-	&	0.58	&	-	 &	 550	&	 40	 &	3.5	&	 \cite{2011AJ....141..180H} 	 &	S & 15	\\
  2007.3286&	  -	   &	-	&	-	    &	-	&	1.63	&	-	&	 698	 &	 40	 &	3.5	&	 \cite{2011AJ....141..180H} 	 &	S	& 0 \\
  2009.2601&	305.5	&	0.7	&	0.0383	&	0.0002&	0.4	&	-	 &	551	 &	 22	&	 4.1	 &	\cite{2010AJ....139..743T}	&	 S  & 15	 \\
  2009.4462&	320.8	&	-	&	0.0318	&	-	&	0.46	&	-	 &	 562	&	 40	 &	3.5	&	 \cite{2011AJ....141..180H}  	 &	S	& 15\\
  2009.4462&	  -	   &	-	&	-	    &	-	&	0.31	&	-	&	 692	 &	 40	 &	3.5	&	 \cite{2011AJ....141..180H} 	 &	S & 0	\\
  2009.4462&	322.6	&	3.2	&	0.044	&	0.003&	0.45	&	 0.12	 &	 562	&	 40   	 &	3.5	&	\cite{2012AJ....143...10H}	&	S & 15	\\
  2009.4462&	322.2	&	3.2	&	0.044	&	0.003&	0.31	&	 0.12	 &	 692	&	 40   	 &	3.5	&	  \cite{2012AJ....143...10H}	&	S &  15	\\
  2012.1843&	8.0	   &	0.0	&	0.1214	&	0.0002&	0.6	   &	-	&	 543	 &	 22	&	 4.1	&	 \cite{2012AJ....144...56T}  	 &	S &	15\\
  2013.4026&  16.5      & -     & 0.1258    & -     & 0.36      & -     &  692      & 40     & 3.5  & Horch**   & S     & 15 \\
  2013.4026&  16.7      & -     & 0.1264    & -     & 0.35      & -     & 880       & 50     & 3.5  & Horch**    & S     & 15 \\
  2014.3005&  23.0      & 0.0   & 0.1204    & 0.0002  & 0.6     & -  & 534  & 22    & 4.2   &  \cite{2015AJ....150...50T}& St & 15\\
\end{longtable}
\flushleft{* References as listed in the Fourth Catalog of Interferometric Measurements of Binary Stars
 \cite{2001AJ....122.3480H} (online version).\\
** Unpublished measurements}
\end{center}

The first column in Table~\ref{deltam2} indicates the date of observation. Columns 2, 3, 4, and 5 give the values of the position angle and the separation with their corresponding standard errors. Columns 6 and 7 show the observed difference of magnitude between the components, while columns 8 and 9 include the wavelength and the standard error used in the observation. The size of the telescope utilized to perform the measurements is indicated in column 10. Finally, columns 11, 12, and 13 contain the reference of the publication where the measurements were announced, the technique used, and the weight assigned to the observation, respectively.

The orbital elements and the masses determined in the present work together with their estimated standard errors are included in Table~\ref{orbitalelements}.

Table~\ref{rms} lists the rms residuals and mean arithmetic residuals of the position angles and separations for the orbits calculated in this work as well as for previously determined orbits. Table~\ref{ephemerides} presents the ephemerides for each orbit for the period between 2014 and 2020. Keeping in mind the ephemerides of both solutions, it will be possible to discriminate between the two calculated orbits in the very near future.

The similarity between the value of the Hipparcos parallax ($12.28\pm0.77$ mas) with the dynamical parallaxes obtained for each orbit (short period: $13.12\pm0.31$ mas, and long period: $13.38\pm0.46$ mas), demonstrates the robustness of the first as well as its use as a referent value in this work.

We concluded that the true orbit is that of the short-period because the long-period orbit gives unacceptable residuals in the position angle regarding the observations of 1990.2759 and 2007.0105. Even if we give 0 weight to those observations, the rms in the position angle is worse in the long-period orbit (see Table~\ref{resid}).

\begin{figure}
%\resizebox{0.6\hsize}{!} {
\includegraphics[width=0.4\textwidth]{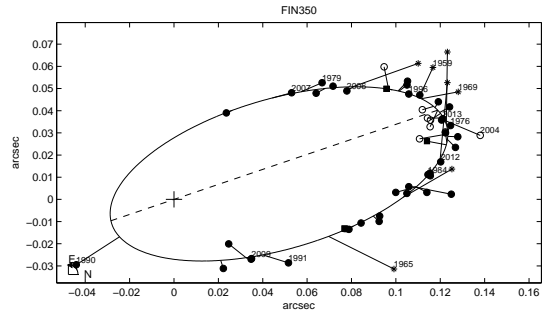}
 \label{orbit9}
  \caption{The apparent short period orbit (P$\sim$9 Yr). Stars represent the measurements made by Finsen; open circles, dots, and rectangles are the measurements carried out with 1-2, 3-4, and 6 m telescope class respectively. Dates of several observations (rounded to the nearest integer) are included.}
\end{figure}

\begin{figure}
%\resizebox{0.6\hsize}{!} {
\includegraphics[width=0.4\textwidth]{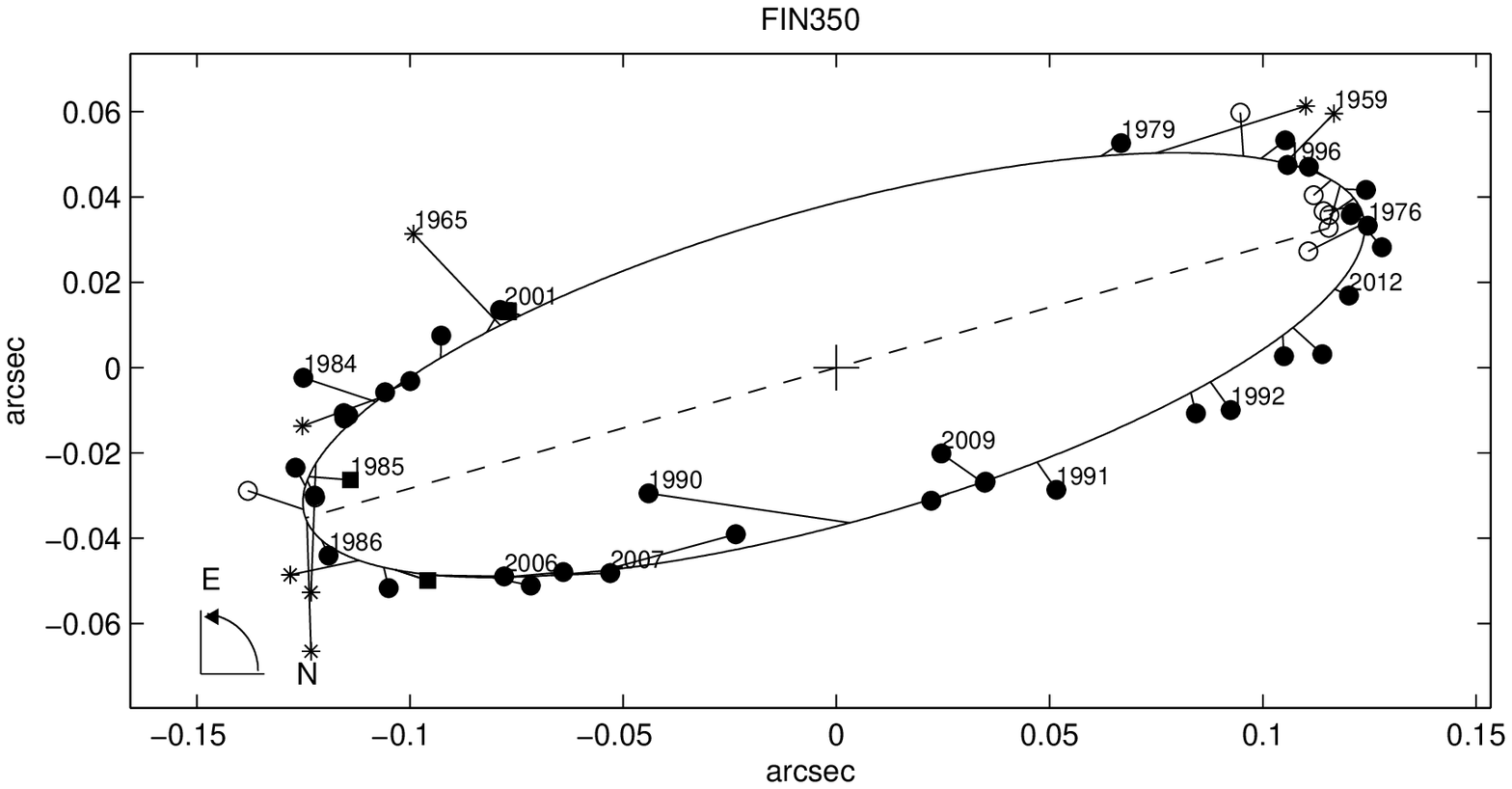}
 \label{orbit18}
  \caption{The apparent long period orbit (P$\sim$18 Yr). Stars represent the measurements made by Finsen; open circles, dots, and rectangles are the measurements carried out with 1-2, 3-4, and 6 m telescope class respectively. Dates of several observations (rounded to the nearest integer) are included.}
\end{figure}

%\clearpage

%\begin{deluxetable}{lcccccccccccc}
%\tabletypesize{\scriptsize}
%%\rotate
%%\tablewidth{0pt}
%\label{deltam2}
%\tablecaption{Observational data of Finsen 350}
%\tablehead{
%\colhead{Date}           & \colhead{$\theta(deg)$}      &
%\colhead{$\sigma\theta(deg)$}          & \colhead{$\rho(arcsec)$}  &
%\colhead{$\sigma\rho(arcsec)$}          & \colhead{$\delta m$}    &
%\colhead{$\sigma\delta m$}  & \colhead{Filter $\lambda$}  &
%\colhead{$\delta\lambda$} & \colhead{Tel.} &
%\colhead{Ref.*} &   \colhead{Meth}  &   \colhead{weight}}
%\startdata
%\enddata
%\tablenotetext{*}{References are abbreviated as in the Fourth Catalog of Interferometric Measurements of Binary Stars \citep{2001AJ....122.3480H} (online version)}
%\tablenotetext{**}{Unpublished measurements}
%%% You can append references to a table using the \tablerefs command.
%\end{deluxetable}

%*******************************************************************************
\begin{table*}
\begin{center}
\caption{Calculated orbital elements and masses of the system with standard errors}% \small
 \label{orbitalelements}
\begin{tabular}{@{}lcc@{}}
\hline \hline
 Parameters & Short period & Long period \\
  \hline
 Period, $P $ & 9.130 yr $\pm$  0.030&   18.442 yr $\pm$ 0.200\\
  Periastron epoch, $T_0$ & 2017.487   $\pm$   0.050&   2009.328    $\pm$  0.900\\
  Eccentricity, $e$  &  0.622 $\pm$  0.007&  0.021  $\pm$  0.007\\
Semi-major axis, $a$ & 0''0795 $\pm$  0.002&  0''129 $\pm$   0.002\\
  Inclination, $i$ & 57$^{\circ}$.0  $\pm$  0.5&  73$^{\circ}$.5   $\pm$   0.5\\
Position angle of nodes, $\Omega$&  18$^{\circ}$.8 $\pm$  1.5&  15$^{\circ}$.8 $\pm$   1.0\\
 Argument of periastron, $\omega$&  170$^{\circ}$.8  $\pm$  3.5&   279$^{\circ}$.3   $\pm$ 18.0\\
Mass sum$^{*}$ $\Sigma M_{A,B}(M_{\odot})$ &3.255$\pm$0.332&3.408$\pm$0.311\\
Mass sum$^{**}$ $\Sigma M_{A,B}(M_{\odot})$ &3.173$\pm$0.152&3.146$\pm$0.149\\
Mass sum$^{***}$ $\Sigma M_{A,B}(M_{\odot})$ &2.665$\pm$0.125&2.637$\pm$0.122\\
   \hline \hline
\end{tabular}
\end{center}
$^{*}$ Using Hipparcos parallax: 12.28 mas.\\
$^{**}$ Using dynamical parallaxes: $12.39\pm 0.31$ mas (short period) and $12.61\pm 0.36$ mas (long period), and the calibration for main sequence stars.\\
$^{***}$ Using dynamical parallaxes: $13.13\pm 0.43$ mas (short period) and $13.38\pm 0.46$ mas (long period), and the calibration for subgiants given in \citep{docobo2013dynamical}.
\end{table*}

\begin{table}
\begin{center}
\caption{RMS and Mean Arithmetic Residuals for the new and old orbits}
 \label{rms}
\begin{tabular}{@{}lcccc}
\hline \hline
 & \multicolumn{2}{c}{RMS} & \multicolumn{2}{c}{Mean Arithmetic} \\
  \hline
  Authors & $\Delta \theta$ & $\Delta \rho$ & $\Delta \theta$ & $\Delta \rho$ \\
  \hline
This paper, 9 yr. & 2.208 & 0.008 & 0.003 & 0.001 \\
This paper, 18 yr. & 9.289 & 0.009 & -0.938 & -0.000 \\
Baize & 12.136 & 0.011 & -0.085 & 0.004 \\
Hartkopf & 6.191 & 0.009 & -2.062 & -0.001 \\
Horch & 2.673 & 0.008 & -0.031 & -0.001 \\
   \hline \hline
\end{tabular}
\end{center}
\end{table}

\begin{table*}
\begin{center}
\small
\caption{Residuals of the orbits.} \label{resid}
\begin{tabular}{lrrrr|lrrrr}
\hline \hline \\[-2ex]

& \multicolumn{2}{c}{Long period}   &   \multicolumn{2}{c}{Short period}& &  \multicolumn{2}{c}{Long period}
   &   \multicolumn{2}{c}{Short period}
             \\[0.5ex] \hline

     \multicolumn{1}{c}{Date}   &   \multicolumn{1}{c}{$\Delta \theta(^{\circ})$}&   \multicolumn{1}{c}{$\Delta \rho('')$}
   &   \multicolumn{1}{c}{$\Delta \theta(^{\circ})$}&   \multicolumn{1}{c}{$\Delta \rho('')$}
   &   \multicolumn{1}{c}{Date}   &   \multicolumn{1}{c}{$\Delta \theta(^{\circ})$}&   \multicolumn{1}{c}{$\Delta \rho('')$}
   &   \multicolumn{1}{c}{$\Delta \theta(^{\circ})$}&   \multicolumn{1}{c}{$\Delta \rho('')$}
       \\[0.5ex] \hline
   \\[-1.8ex]

1959.47 & 2.5 & 0.015 & 4.4 & 0.011 &1987.3800 & 2.9 & -0.006 & 0.7 & -0.003 \\
1960.55 & -4.8 & 0.036 & -3.2 & 0.030 &1988.1655 & 4.2 & -0.007 & 0.3 & -0.000 \\
1964.530 & -10.4 & 0.025 & -4.2 & 0.032 &1988.2524 & 4.6 & -0.012 & 0.4 & -0.005\\
1965.545 & 2.5 & 0.019 & 5.6 & 0.022 &1990.2759 & -61.2 & 0.017 & -0.8 & 0.023\\
1966.520 & 11.1 & -0.124 & 12.8 & -0.122 &1991.3186 & -3.8 & 0.007 & 2.8 & 0.011\\
1966.527 & 12.7 & 0.010 & 14.4 & 0.012 &1992.3098 & -3.2 & 0.002 & -0.5 & -0.001\\
1967.5447 & 12.2 & 0.011 & 12.8 & 0.012 &1992.4572 & -4.0 & 0.005 & -1.4 & 0.002\\
1968.545 & -1.2 & 0.016 & -1.5 & 0.016 &1993.0905 & -2.7 & 0.000 & -0.6 & -0.003\\
1976.2959 & -2.4 & 0.003 & -0.6 & 0.004 &1993.1973 & -3.4 & 0.007 & -1.5 & 0.004\\
1976.3697 & -1.5 & -0.014 & 0.3 & -0.014 &1995.1495 & 0.6 & -0.008 & 0.9 & -0.008\\
1976.4570 & -0.8 & 0.000 & 0.8 & 0.001 &1995.3109 & -0.9 & -0.007 & -0.8 & -0.006\\
1977.0877 & -0.8 & 0.005 & 0.3 & 0.004 &1996.1840 & 0.8 & -0.003 & 0.1 & -0.001\\
1977.1751 & -4.0 & -0.006 & -2.9 & -0.006 &2001.2708 & -0.3 & 0.003 & -0.4 & -0.002\\
1977.3280 & -0.9 & -0.005 & 0.0 & -0.006 &2001.2708 & -0.4 & 0.004 & -0.5 & -0.001\\
1978.1499 & 0.6 & 0.007 & 0.9 & 0.005 &2001.2708 & -0.3 & 0.003 & -0.4 & -0.002\\
1978.3109 & 4.8 & 0.004 & 4.9 & 0.003 &2002.3224 & -1.0 & -0.004 & -1.2 & -0.010\\
1979.3622 & -0.4 & 0.006 & -2.1 & 0.009 &2002.3224 & 0.3 & 0.002 & 0.1 & -0.004\\
1983.0701 & -3.9 & -0.002 & -1.4 & -0.002 &2004.1960 & -3.1 & 0.012 & -4.6 & 0.013\\
1983.4332 & -3.2 & 0.000 & -1.3 & -0.001 &2006.1915 & 4.7 & -0.014 & -0.1 & -0.004\\
1984.0532 & -3.2 & 0.017 & -1.9 & 0.015 &2007.0105 & 6.2 & -0.011 & -3.8 & 0.007\\
1984.3752 & -1.0 & 0.000 & -0.1 & -0.001 &2007.3286 & 18.0 & -0.027 & 2.0 & -0.003\\
1984.3807 & -1.4 & 0.001 & -0.4 & -0.000 &2009.2601 & -6.4 & -0.001 & -2.0 & 0.003 \\
1984.3835 & -0.8 & 0.001 & 0.2 & -0.000 &2009.4462 & -1.1 & -0.012 & -0.3 & -0.010\\
1985.1805 & 1.3 & -0.009 & 1.5 & -0.009 &2009.4462 & 0.7 & 0.001 & 1.5 & 0.002\\
1985.2438 & 1.9 & -0.001 & 2.1 & -0.000 &2009.4462 & 0.3 & 0.001 & 1.1 & 0.002 \\
1985.3389 & -2.1 & 0.001 & -2.0 & 0.002 &2012.1843 & -0.9 & 0.003 & -0.8 & -0.001\\
1985.4840 & 0.4 & -0.002 & 0.3 & -0.002 &2013.4026 & 0.4 & -0.003 & -0.4 & -0.002\\
1986.4067 & 1.8 & -0.000 & 0.8 & 0.001 &2013.4026 & 0.6 & -0.002 & -0.2 & -0.001 \\
1987.2642 & 2.4 & 0.001 & 0.4 & 0.004 &2014.3005 & 1.7 & -0.003 & -0.0 & 0.001\\
1987.3800 & 2.9 & -0.006 & 0.7 & -0.003 & &&&&\\

\hline
\end{tabular}
\end{center}
\end{table*}

\begin{table}
\begin{center}
\caption{Ephemerides for calculated orbits from 2016 to 2020}
 \label{ephemerides}
\begin{tabular}{@{}lcccc}
\hline \hline
 & \multicolumn{2}{c}{Short period} & \multicolumn{2}{c}{Long period} \\
  \hline
  t & $\theta$ & $\rho$ & $\theta$ & $\rho$ \\
  \hline
2016.0 & 42.6 & 0.070 & 35.4 & 0.087 \\
2017.0 & 110.6 & 0.024 & 54.1 & 0.057 \\
2018.0 & 259.6 & 0.027 & 102.3 & 0.037 \\
2019.0 & 338.9 & 0.059 & 154.2 & 0.054 \\
2020.0 & 357.1 & 0.095 & 174.8 & 0.084 \\
   \hline \hline
\end{tabular}
\end{center}
\end{table}

\section{Atmospheric Modeling}
\label{Atm}
In order to estimate the physical and geometrical parameters of the individual components of the system, we follow Al-Wardat's complex method for analyzing CVBS~\citep{2012PASA...29..523A}. The method makes use of the measured magnitude difference $\Delta m$ between the sub-components, their composite visual magnitude $m_v$ and the parallax of the system to calculate preliminary input parameters to model atmospheres of the individual components. Model atmospheres are then used to calculate their spectral energy distributions (SEDs), which then combined together (according to specific criteria) to build the entire synthetic SED of the system. The observational SED is used as a reference guide to the synthetic one in an iterated way of the aforementioned steps by changing the input parameters until the best fit between them achieved.

The magnitude difference between the two components  $\Delta m=m_B-m_A = 0^m.59$ was taken as the average value of all measurements under the filters 550/40, 551/22, 562/40, 543/22 and 534/22 (given in Table~\ref{deltam2}), which are the closest to the V-band filter.   This magnitude difference, along with the composite photometry of the system $m_v=6^m.358$ (Table~\ref{stardata}), was used as an input to the equations:

\begin{eqnarray}\label{Ma}
m_A=m_v+2.5log(1+10^{-0.4\Delta m}),\\
 m_B=m_A+\Delta m,
\end{eqnarray}

\noindent  to calculate the apparent magnitudes of the individual components as: $m_A=6^m.87$ and $m_B=7^m.42$.

These individual apparent magnitudes, along with the corresponding main sequence relations and standard values \citep{2005oasp.book.....G, 1992adps.book.....L}.

\begin{eqnarray}\label{equ2}
M_v=m_v+5-5\log(d)-A_\nu,\\
\log(R/R_\odot)= 0.5 \log(L/L_\odot)-2\log(T/T_\odot),\ and\\
\log g = \log(M/M_\odot)- 2\log(R/R_\odot) + 4.43,
\end{eqnarray}

 \noindent were used to calculate the preliminary input parameters  (effective temperatures and surface gravity accelerations) needed to build model atmospheres for the individual components .
 We used bolometric corrections of \cite{2005oasp.book.....G} as well as $T_\odot=5777K$  and extinction ($A_v$) given in Table~\ref{stardata} by NASA/IPAC.\\
Hence, the calculated  parameters were used as input parameters to construction  model atmospheres for each component using grids of Kurucz's 1994 blanketed models (ATLAS9) \citep{1994KurCD..19.....K}. Here, we used solar-abundance line-blanketed model atmospheres to build the  spectral energy distribution for each component.

 The total energy flux received from the binary star is calculated depending on the net luminosity of the components, A and B, located at a distance, d, from the Earth. This is represented by the following equation ~\citep{2002BSAO...53...51A}:

\begin{eqnarray}\label{F}
F_{\lambda} \cdot d^2 = H_{\lambda}^{\rm A} \cdot R_{\rm A}^2 +
H_{\lambda}^{\rm B} \cdot R_{\rm B}^2\,
\end{eqnarray}

\noindent where $H_{\lambda}^{\rm A}$ and $H_{\lambda}^{\rm B}$ are the fluxes from a unit surface of the corresponding component and $F_{\lambda}$  represents the total SED of the system.

\begin{figure}%[!h]
\resizebox{0.4\hsize}{!} {\includegraphics[]{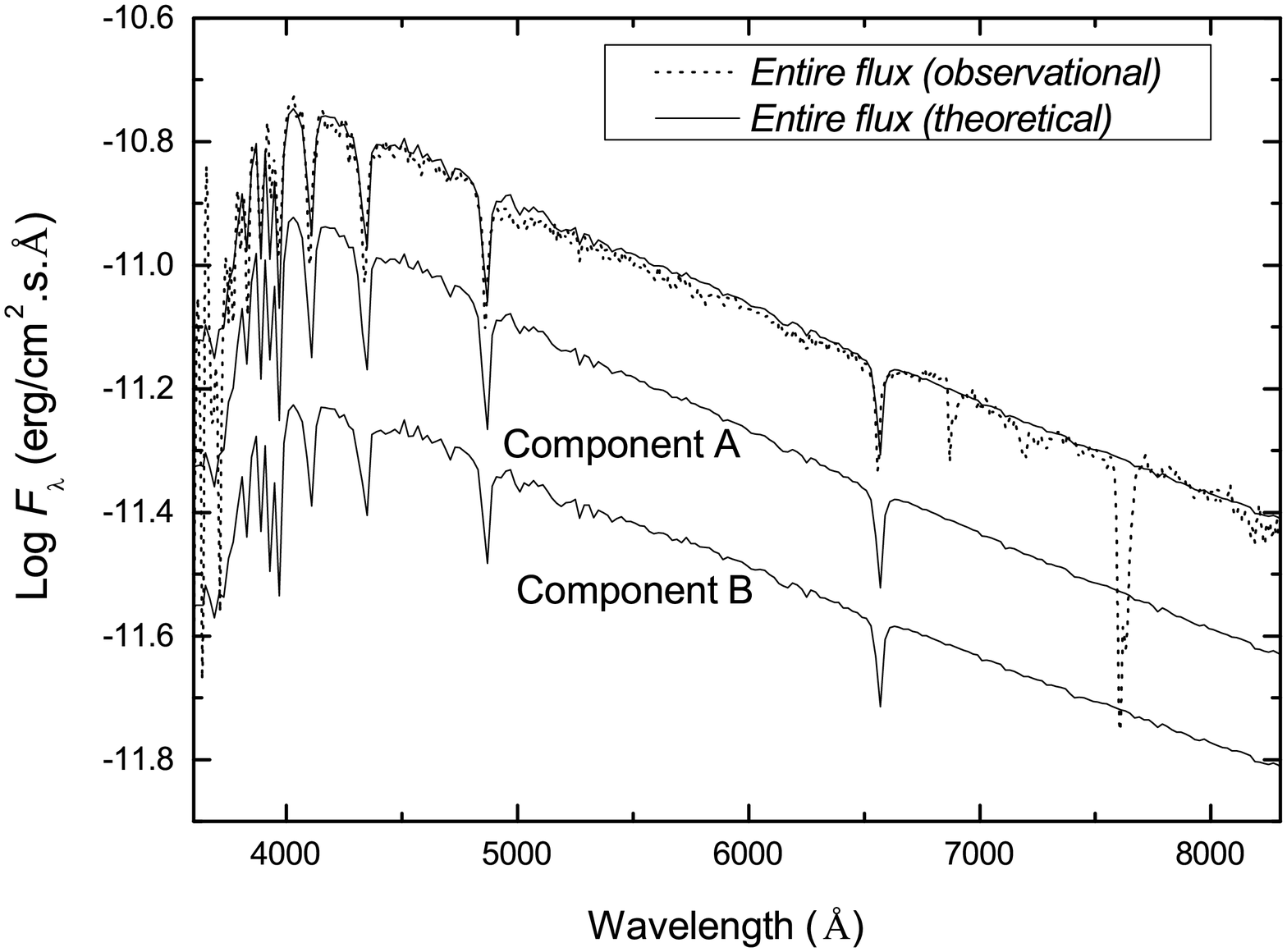}}
 \caption{Best fit between the entire observational SED of the system taken from ~\citep{2002BSAO...53...58A} and the entire synthetic one of the two components built in this work. The figure also shows  the SEDs of the individual components built using the parameters given at the end of section ~\ref{Atm}.}
 \label{logspectral}
\end{figure}

Now, the goal is to achieve the best fit between the computed total  SED with the observed one. So, in order to achieve that fit, dozens of different sets of parameters were tested by different ways; the first way is the  direct correspondence  as can be seen in Fig.\ref{logspectral},  which includes the maximum values of the absolute flux, the shape of the continuum, and the profiles of the absorption lines. The second way is by comparing  synthetic magnitudes and color indices with the observational ones (see Table ~\ref{tablecopmarison}).

It is worthwhile to mention here that two of the input parameters have the same effect on the maximum values of the absolute flux, these are the radii of the components and the parallax of the system according to Equ. \ref{F}. Hence, the radii of both components were set subject to change according  to  the parallaxes of different sources. Table ~\ref{radii} gives these radii using  the following atmospheric parameters:
\begin{eqnarray}\label{equ4}
\nonumber
T_{\rm eff}^{\rm A} =7820\pm75\,{\rm K},\ T_{\rm eff}^{\rm B} =7250\pm75\,{\rm K},\\
\nonumber
\log g_{\rm A}=4.10\pm0.40,\ \log g_{\rm B}=4.25\pm0.40
\end{eqnarray}

 %and changing the radii until obtain the best absolute flux, keeping in mind the value of $\Delta m$. Then we tested the parallax by fixing the radii as given by Tables B.1 and B.2 in \citet{2005oasp.book.....G} or the standard T-L-R equation for the main sequence stars of $T^A_{eff}=7600 K$, $T^B_{eff}=7200 K$, and modifying the parallax until the best absolute flux is determined.

%*******************************************************************************
\begin{table*}
\begin{center}
\caption{Estimated radii and luminosities of the individual components according to different parallaxes.}% \small
 \label{radii}
\begin{tabular}{@{}lccccc@{}}
\hline \hline
 Source of Parallax  &  $\pi$ (mas) & $R_a \pm0.07$ & $R_b \pm0.07$ &$L_a/L_{\odot}$& $L_b/L_{\odot}$\\
  \hline
Hipparcos  (new)$^{*}$        & 12.28 & 1.92  & 1.71 & 12.38 &7.25\\
 Dynamical parallax & 12.39 & 1.88  & 1.67  & 11.87 & 6.92\\
 (short period, MS) & &   &  && \\
 Dynamical parallax & 12.61 & 1.86  & 1.65 & 11.62 & 6.75 \\
(long period, MS) &  &   &   &&\\
 Dynamical parallax & 13.13 & 1.79  & 1.60  & 10.76 & 6.35\\
 (short period, Subgiant) & &   &  && \\
 Dynamical parallax & 13.38 & 1.76  & 1.57 & 10.40 & 6.11 \\
(long period, Subgiant) &  &   &   &&\\
 Hipparcos  (old)        & 13.45 & 1.75  & 1.56 & 10.28 & 6.04\\

   \hline \hline
\end{tabular}
\end{center}
$^{*}$ \cite{2007A&A...474..653V}
\end{table*}

The question now is which parallax represents better the system?. It is clear from Table  ~\ref{radii} that  Hipparcos old parallax gives the highest radii and luminosities, which means if this parallax is true, then the components are at the beginning of the subgiant phase.
Comparing the masses estimated by orbital analysis ($\Sigma M_{A,B}(M_{\odot})$=3.173$\pm$0.152 for a short-period main sequence pair and $\Sigma M_{A,B}(M_{\odot})$ =2.665$\pm$0.125 for a subgiant pair)(Table ~\ref{orbitalelements}) with the positions of both components on the evolutionary tracks (Fig. ~\ref{evol}), we find that the short period solution with its dynamical parallax is the best solution for this system.

So, the best fit between the observed and synthetic entire spectra as represented in Fig. ~\ref{logspectral}  was achieved using the parameters listed in Table ~\ref{tablef1} with the dynamical parallax $\pi$ (mas)= 12.39$\pm$0.31.
%\begin{eqnarray}
%\nonumber
%T_{\rm eff}^{\rm A} =7820\pm75\,{\rm K},\ T_{\rm eff}^{\rm B} =7250\pm75\,{\rm K},\\
%\nonumber
%\log g_{\rm A}=4.10\pm0.40,\ \log g_{\rm B}=4.25\pm0.40\,\\
%\nonumber
%R_{\rm A}=1.88\pm0.07\, R_{\rm B}=1.67\pm0.07\,\\
%\nonumber
%\pi (mas)= 12.39\pm0.31
%\end{eqnarray}

%Hence, the spectral types follow as $Sp^A=A7$ and $Sp^B=F0$.

%Fig. ~\ref{logspectral} shows the best fit between the entire observational SED of the system taken from  ~\citep{2002BSAO...53...58A} and the entire synthetic SED computed using Al-Wardat's complex method based on Kurucz blanketed models \citep{1994KurCD..19.....K} and using Hipparcos parallax. The figure also shows  the SEDs of the individual components built using the following parameters:   $T^A_{eff}=7600\pm75K$, $log g_A=4.08\pm0.40$, $R_A=1.90\pm0.12R_\odot$,   $T^B_{eff}=7200\pm75K$, $log g_B=4.08\pm0.40$, and $R_B=1.76\pm0.11R_\odot$.

\subsection{Synthetic photometry}

In addition to their importance as descriptive parameters, apparent magnitudes and color indices of the individual components and the entire system play an important role in testing the best fit between the synthetic and observational SEDs  as mentioned in paragraph 6 of Sec. ~\ref{Atm}.

In order to calculate the synthetic magnitudes,  we used the following relation \citep{apellaniz2006recalibration} and \citep{maiz2007uniform}:

\begin{eqnarray}\label{equ5}
%\nonumber
m_p[F_{\lambda,s}(\lambda)] = -2.5 \log \frac{\int P_{p}(\lambda)F_{\lambda,s}(\lambda)\lambda{\rm d}\lambda}{\int P_{p}(\lambda)F_{\lambda,r}(\lambda)\lambda{\rm d}\lambda}+ {\rm ZP}_p,
\end{eqnarray}

\noindent  where $m_p$ is the synthetic magnitude of the passband p,  $P_p(\lambda)$ is the dimensionless sensitivity function of the passband p, $F_{\lambda,r}(\lambda)$ is the synthetic SED of the object and  is the SED of the reference star (Vega). Zero points ($ZP_p$) from \cite{maiz2007uniform} (and references therein) were adopted.

The results of the calculated magnitudes and color indices of the combined system and individual components in different photometric systems are given in Table~\ref{sed-result}:

\begin{table}[!ht]
\small
\begin{center}
\caption{Magnitudes and color indices of the synthetic spectra of the system}
\label{sed-result}
\begin{tabular}{lcccc}
\hline\hline
System. & Filter & Entire & Comp. A  & Comp. B   \\
\hline
Johson-        & $U$ & 6.67   & 7.15 & 7.78 \\
\, Cousins     & $B$ & 6.61   &  7.07 &7.76  \\
               & $V$ & 6.36  &  6.85 &  7.44 \\
               & $R$ & 6.22   &  6.74 & 7.27  \\
               &$U-B$& 0.06   & 0.08 & 0.02 \\
               &$B-V$&0.25   &  0.22 & 0.32 \\
               &$V-R$& 0.14   &  0.12 & 0.18 \\
  \hline
Str\"{o}mgren       & $u$ & 7.95 & 8.45 &  9.02  \\
                    & $v$ & 6.82  & 7.28  & 8.00 \\
                    & $b$ & 6.49 & 6.97 &  7.62 \\
                    &  $y$& 6.34 & 6.83 &  7.42  \\
                    &$u-v$& 1.12 & 1.18&  1.02 \\
                    &$v-b$& 0.33& 0.31 & 0.37 \\
                    &$b-y$& 0.16& 0.13& 0.20 \\
  \hline
  Tycho       &$B_T$  & 6.68 & 7.14 & 7.84   \\
              &$V_T$  & 6.39 & 6.89 & 7.49  \\
              &$B_T-V_T$& 0.29& 0.25& 0.35\\
\hline \hline
\end{tabular}
\end{center}
\end{table}

\section{Results and discussion}

Tables \ref{orbitalelements},\ref{sed-result} and \ref{tablef1} list the estimated physical and geometrical parameters of the system, which represent adequately enough the system.  Fig. ~\ref{logspectral} shows a good consistency between the synthetic and observational SEDs.

The orbital analysis of the system gives two solutions;  the 9.130 yr short period orbit with mass sum  3.255 $M_{\odot}$ and the 18.442 yr long period orbit with mass sum 3.408 $M_{\odot}$ using 12.28 mas Hipparcos parallax. For comparison,
\cite{1998A&A...330..585M} found that $\Sigma M=2.869 M_{\odot}\pm 0.684, M_A=1.616M_{\odot}\pm 0.422$ and $ M_B=1.253M_{\odot}\pm 0.345$ using $\pi=12.92\pm 0.95$mas and $\Delta m=0.010\pm 0.15$.

The estimated mass sum using atmospheric analysis ($\Sigma M=3.30 M_{\odot}, M_A=1.75M_{\odot}$ and $ M_B=1.55M_{\odot}$, Table~\ref{tablef1}) supports that of the  short period orbit solution as 3.255 $M_{\odot}$.

The comparison between the observational and synthetic magnitudes of the entire system (Table ~\ref{tablecopmarison}) gives a good indication about the reliability of Al-Wardat's complex method in analyzing CVBS.

\begin{figure}[!ht]
\resizebox{0.6\hsize}{!} {\includegraphics[]{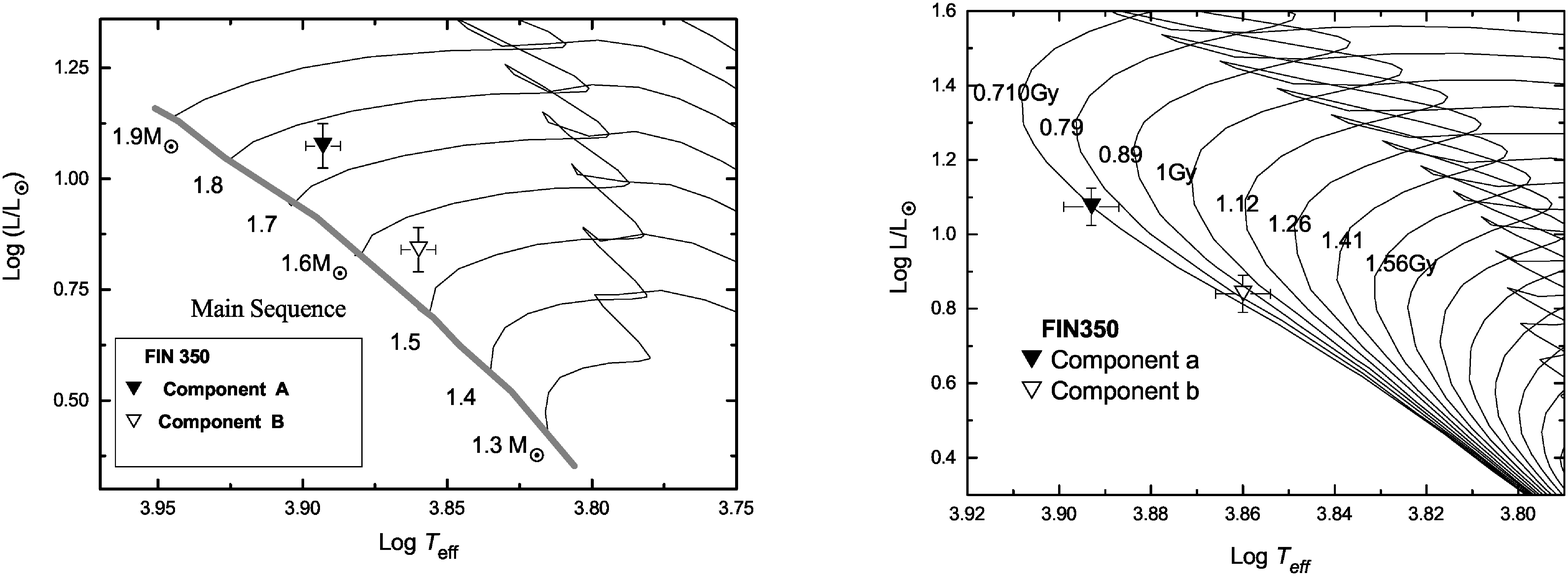}}
 \caption{The systems' components on the evolutionary tracks for low- and intermediate-mass stars (left) and on the isochrones of low- and intermediate-mass, solar composition [$Z$=0.019, $Y$=0.273] stars (right) of ~\cite{2000A&AS..141..371G}.}
 \label{evol}
\end{figure}

 Figs. ~\ref{evol} and ~\ref{isoch}  give the positions of the two components on the evolutionary tracks and isochrones for low- and intermediate-mass stars of ~\cite{2000A&AS..141..371G}; the error bars in the figure represent the effect of the parallax and radii uncertainty.

 %Component A has a higher effective temperature, mass and radius, and  shows an advanced position on the evolutionary tracks with an estimated age of 3.54 Gy and A7IV spectral type and luminosity class, while component B has a lower effective temperature, mass and radius, with an estimated age of 5.5 Gy and F0IV spectral type and luminosity class.

 %It is worthwhile to mention here that \cite{1975AJ.....80..637M} assigned the system to the F0V MK spectral type,  while \cite{1976PASP...88...95C} assigned it to A7IV (or mild A7m).

 Depending on the estimated parameters of the system's components and their positions on the evolutionary tracks with age around 0.79 Gy (Fig. ~\ref{evol}),   fragmentation is the most likely process for the formation of such system. Where \cite{1994MNRAS.269..837B} concluded that fragmentation of rotating disk around an incipient central protostar is possible, as long as there is continuing infall, and \cite{2001IAUS..200.....Z} pointed out that hierarchical  fragmentation during rotational collapse has been invoked to produce binaries and multiple systems.

\section{Conclusions}
Al-Wardat's complex method along with Docobo's analytical method for orbit calculation were used to analyze the speckle interferometric close visual binary star FIN 350 (WDS 13175-0041, HIP 64838, HD 115488). The physical and geometrical parameters of the system's components were estimated depending on the orbital solution of the system and  the best fit between the entire observational SED and the synthetic ones built using model atmospheres.

The dynamical parallax ($\pi$=12.39$\pm$0.31, which lies between the old and new Hipparcos measurements) gives the best coincidence between Al-Wardat's complex analysis and Docobo's analytical solution for this system, and it was adopted as the parallax of the system.

New orbits (short and long-period) of the system  were calculated. The 9.130 yr  short-period improves the earlier orbits while the 18.442 yr long-period solution was calculated for the first time. Nevertheless, in this work it was demonstrated that the short-period is the orbit that better fits the observations.

The synthetic magnitudes and colors of the entire system and individual components were computed in different photometric systems as given in Table ~\ref{sed-result}. In addition to their importance as parameters, these synthetic magnitudes and colors show the accuracy of the method.

The spectral types and luminosity classes of the components of the system were concluded as A7V for the component A and F0V for the component B, which assured the main sequence phase of both components.

%Depending on the estimated parameters of the system's components and their positions on the evolutionary tracks,  fragmentation process was proposed as the most likely process for the formation of the system.

\begin{table}
\small
\begin{center}
\caption{Comparison between the observational and synthetic magnitudes of the entire system.}
\label{tablecopmarison}
\begin{tabular}{lcc}
\hline \hline
  & $\textrm{Obs}$    & Synth. (this work)  \\
\hline
\noalign{\smallskip}
  $V_J$     & 6.36 & 6.36 \\
  $B_T$     & 6.66 & 6.68 \\
  $V_T$     & 6.38 & 6.40 \\
  $(B-V)_J$ & 0.26 & 0.252\\
  $\triangle m $& 0.543& 0.548\\
  $b-y$   & 0.164* & 0.16 \\
\hline \hline
\end{tabular}
\end{center}
* \cite{1972A&A....18..428D}
\end{table}

\begin{table}
\begin{center}
\caption{Parameters for the components of the system}
\label{tablef1}
\begin{tabular}{lcc}
\hline
Component & A &  B  \\
\hline
\noalign{\smallskip}
$T_{\rm eff}$\,(K) & $7820\pm75$ & $7250\pm75$ \\
Radius (R$_{\odot}$) & $1.88\pm0.07$ & $1.67\pm0.07$ \\
$\log g$ & $4.10\pm0.40$ & $4.25\pm0.40$ \\
$L (L_\odot)$ & $11.87\pm1.20$  & $6.92\pm0.70$\\
$M_{V}$ & $2.19\pm0.26$& $2.78\pm0.30$\\
Mass, ($M_{\odot})$*& $1.75\pm0.18$ & $1.55\pm0.16$  \\
Sp. Type** & A7V & F0V \\
\hline
\end{tabular}
\end{center}
* Depending on the positions of the components on the evolutionary tracks of \cite{2000A&AS..141..371G}.\\
** Depending on the Tables of \cite{1970A&A.....4..234F}; \cite{2005oasp.book.....G}.
\end{table}

\section*{Acknowledgments}
%\begin{acknowledgements}
The authors thank Elliott Horch for providing two unpublished speckle measurements used to confirm the orbit calculated. They also thank Suhail Masda from Astrophysikalisches Institut und Universit\"{a}ts-Sternwarte Jena, Germany for his help in some calculations, the anonymous referee(s) for the valuable comments and Ms. Asmaa Ramadan for her help in the language editing.
Also,
This research was financed by the AYA2011-26429 project funded by the Spanish Ministerio de Econom\'ia y Competitividad.
In this work, the authors made use of the Fourth Interferometric Catalogue \citep{2001AJ....122.3480H}, SIMBAD database and CHORIZOS code of photometric and spectrophotometric data analysis \citep{2004PASP..116..859M}.
%\end{acknowledgements}

\newpage
%\bibliographystyle{mn2e}
%\bibliography{alwardat2}

\begin{thebibliography}{}

\bibitem[\protect\citeauthoryear{{Al-Wardat}}{{Al-Wardat}}{2002a}]{2002BSAO...53...51A}
{Al-Wardat} M.~A.,  2002a, Bull.~Special Astrophys.~Obs., 53, 51

\bibitem[\protect\citeauthoryear{{Al-Wardat}}{{Al-Wardat}}{2002b}]{2002BSAO...53...58A}
{Al-Wardat} M.~A.,  2002b, Bull.~Special Astrophys.~Obs., 53, 58

\bibitem[\protect\citeauthoryear{{Al-Wardat}}{{Al-Wardat}}{2007}]{2007AN....328...63A}
{Al-Wardat} M.~A.,  2007, Astronomische Nachrichten, 328, 63

\bibitem[\protect\citeauthoryear{{Al-Wardat}}{{Al-Wardat}}{2009}]{2009AN....330..385A}
{Al-Wardat} M.~A.,  2009, Astronomische Nachrichten, 330, 385

\bibitem[\protect\citeauthoryear{{Al-Wardat}}{{Al-Wardat}}{2012}]{2012PASA...29..523A}
{Al-Wardat} M.~A.,  2012, \pasa, 29, 523

\bibitem[\protect\citeauthoryear{{Al-Wardat}}{{Al-Wardat}}{2014}]{2014AstBu..69..454A}
{Al-Wardat} M.~A.,  2014, Astrophysical Bulletin, 69, 454

\bibitem[\protect\citeauthoryear{{Al-Wardat}, {Balega}, {Leushin}, {Yusuf},
  {Taani}, {Al-Waqfi} \& {Masda}}{{Al-Wardat}
  et~al.}{2014}]{2014AstBu..69...58A}
{Al-Wardat} M.~A.,  {Balega} Y.~Y.,  {Leushin} V.~V.,  {Yusuf} N.~A.,  {Taani}
  A.~A.,  {Al-Waqfi} K.~S.,    {Masda} S.,  2014, Astrophysical Bulletin, 69,
  58

\bibitem[\protect\citeauthoryear{{Al-Wardat} \& {Widyan}}{{Al-Wardat} \&
  {Widyan}}{2009}]{2009AstBu..64..365A}
{Al-Wardat} M.~A.,  {Widyan} H.,  2009, Astrophysical Bulletin, 64, 365

\bibitem[\protect\citeauthoryear{{Al-Wardat}, {Widyan} \&
  {Al-thyabat}}{{Al-Wardat} et~al.}{2014}]{2014PASA...31....5A}
{Al-Wardat} M.~A.,  {Widyan} H.~S.,    {Al-thyabat} A.,  2014, \pasa, 31, 5

\bibitem[\protect\citeauthoryear{Apellaniz}{Apellaniz}{2006}]{apellaniz2006recalibration}
Apellaniz J.~M.,  2006, The Astronomical Journal, 131, 1184

\bibitem[\protect\citeauthoryear{{Baize}}{{Baize}}{1988}]{1988A&AS...74..507B}
{Baize} P.,  1988, \aaps, 74, 507

\bibitem[\protect\citeauthoryear{{Balega}, {Balega}, {Maksimov},
  {Malogolovets}, {Pluzhnik} \& {Shkhagosheva}}{{Balega}
  et~al.}{2006}]{2006BSAO...59...20B}
{Balega} I.~I.,  {Balega} A.~F.,  {Maksimov} E.~V.,  {Malogolovets} E.~A.,
  {Pluzhnik} E.~A.,    {Shkhagosheva} Z.~U.,  2006, Bull.~Special
  Astrophys.~Obs., 59, 20

\bibitem[\protect\citeauthoryear{{Balega} \& {Balega}}{{Balega} \&
  {Balega}}{1987}]{1987PAZh...13..508B}
{Balega} I.~I.,  {Balega} Y.~Y.,  1987, Pisma v Astronomicheskii Zhurnal, 13,
  508

\bibitem[\protect\citeauthoryear{{Balega}, {Balega}, {Hofmann}, {Maksimov},
  {Pluzhnik}, {Schertl}, {Shkhagosheva} \& {Weigelt}}{{Balega}
  et~al.}{2002}]{2002A&A...385...87B}
{Balega} I.~I.,  {Balega} Y.~Y.,  {Hofmann} K.-H.,  {Maksimov} A.~F.,
  {Pluzhnik} E.~A.,  {Schertl} D.,  {Shkhagosheva} Z.~U.,    {Weigelt} G.,
  2002, \aap, 385, 87

\bibitem[\protect\citeauthoryear{{Balega}, {Balega} \& {Vasiuk}}{{Balega}
  et~al.}{1991}]{1991AISAO..31...80B}
{Balega} I.~I.,  {Balega} Y.~Y.,    {Vasiuk} V.~A.,  1991, Astrofizicheskie
  Issledovaniia Izvestiya Spetsial'noj Astrofizicheskoj Observatorii, 31, 80

\bibitem[\protect\citeauthoryear{{Balega}, {Balega} \& {Vasyuk}}{{Balega}
  et~al.}{1989}]{1989AISAO..28..107B}
{Balega} I.~I.,  {Balega} Y.~Y.,    {Vasyuk} V.~A.,  1989, Astrofizicheskie
  Issledovaniia Izvestiya Spetsial'noj Astrofizicheskoj Observatorii, 28, 107

\bibitem[\protect\citeauthoryear{Bonneau, Balega, Blazit, Foy, Vakili \&
  Vidal}{Bonneau et~al.}{1986}]{bonneau1986speckle}
Bonneau D.,  Balega Y.,  Blazit A.,  Foy R.,  Vakili F.,    Vidal J.,  1986,
  Astronomy and Astrophysics Supplement Series, 65, 27

\bibitem[\protect\citeauthoryear{{Bonnell}}{{Bonnell}}{1994}]{1994MNRAS.269..837B}
{Bonnell} I.~A.,  1994, \mnras, 269, 837

\bibitem[\protect\citeauthoryear{{Cowley}}{{Cowley}}{1976}]{1976PASP...88...95C}
{Cowley} A.~P.,  1976, \pasp, 88, 95

\bibitem[\protect\citeauthoryear{{Danziger} \& {Faber}}{{Danziger} \&
  {Faber}}{1972}]{1972A&A....18..428D}
{Danziger} I.~J.,  {Faber} S.~M.,  1972, \aap, 18, 428

\bibitem[\protect\citeauthoryear{Docobo \& Andrade}{Docobo \&
  Andrade}{2013}]{docobo2013dynamical}
Docobo J.,  Andrade M.,  2013, Monthly Notices of the Royal Astronomical
  Society, 428, 321

\bibitem[\protect\citeauthoryear{{Docobo}}{{Docobo}}{1985}]{1985CeMec..36..143D}
{Docobo} J.~A.,  1985, Celestial Mechanics, 36, 143

\bibitem[\protect\citeauthoryear{{Docobo}}{{Docobo}}{2012}]{2012ocpd.conf..119D}
{Docobo} J.~A.,  2012, in {Arenou} F.,  {Hestroffer} D.,  eds, Orbital Couples:
  Pas de Deux in the Solar System and the Milky Way {The use of Docobo's
  analytic method for calculating visual double star orbits}.
pp 119--123

\bibitem[\protect\citeauthoryear{{Docobo}, {Tamazian}, {Balega}, {Blanco},
  {Maximov} \& {Vasyuk}}{{Docobo} et~al.}{2001}]{2001A&A...366..868D}
{Docobo} J.~A.,  {Tamazian} V.~S.,  {Balega} Y.~Y.,  {Blanco} J.,  {Maximov}
  A.~F.,    {Vasyuk} V.~A.,  2001, \aap, 366, 868

\bibitem[\protect\citeauthoryear{{Finsen}}{{Finsen}}{1959}]{1959MNSSA..18...31F}
{Finsen} W.~S.,  1959, Monthly Notes of the Astronomical Society of South
  Africa, 18, 31

\bibitem[\protect\citeauthoryear{{Finsen}}{{Finsen}}{1961}]{1961CiUO..120..367F}
{Finsen} W.~S.,  1961, Circular of the Union Observatory Johannesburg, 120, 367

\bibitem[\protect\citeauthoryear{{Finsen}}{{Finsen}}{1965}]{1965ROCi..124...79F}
{Finsen} W.~S.,  1965, Republic Observatory Johannesburg Circular, 124, 79

\bibitem[\protect\citeauthoryear{{Finsen}}{{Finsen}}{1967}]{1967ROCi..126..138F}
{Finsen} W.~S.,  1967, Republic Observatory Johannesburg Circular, 126, 138

\bibitem[\protect\citeauthoryear{{Finsen}}{{Finsen}}{1969}]{1969ROCi..128..187F}
{Finsen} W.~S.,  1969, Republic Observatory Johannesburg Circular, 128, 187

\bibitem[\protect\citeauthoryear{{Fitzgerald}}{{Fitzgerald}}{1970}]{1970A&A.....4..234F}
{Fitzgerald} M.~P.,  1970, \aap, 4, 234

\bibitem[\protect\citeauthoryear{{Girardi}, {Bressan}, {Bertelli} \&
  {Chiosi}}{{Girardi} et~al.}{2000}]{2000A&AS..141..371G}
{Girardi} L.,  {Bressan} A.,  {Bertelli} G.,    {Chiosi} C.,  2000, \aaps, 141,
  371

\bibitem[\protect\citeauthoryear{{Gray}}{{Gray}}{2005}]{2005oasp.book.....G}
{Gray} D.~F.,  2005, {The Observation and Analysis of Stellar Photospheres}

\bibitem[\protect\citeauthoryear{{Hartkopf}, {Mason} \& {McAlister}}{{Hartkopf}
  et~al.}{1996}]{1996AJ....111..370H}
{Hartkopf} W.~I.,  {Mason} B.~D.,    {McAlister} H.~A.,  1996, \aj, 111, 370

\bibitem[\protect\citeauthoryear{{Hartkopf}, {Mason}, {McAlister}, {Roberts}
  Jr., {Turner}, {ten Brummelaar}, {Prieto}, {Ling} \& {Franz}}{{Hartkopf}
  et~al.}{2000}]{2000AJ....119.3084H}
{Hartkopf} W.~I.,  {Mason} B.~D.,  {McAlister} H.~A.,  {Roberts} Jr. L.~C.,
  {Turner} N.~H.,  {ten Brummelaar} T.~A.,  {Prieto} C.~M.,  {Ling} J.~F.,
  {Franz} O.~G.,  2000, \aj, 119, 3084

\bibitem[\protect\citeauthoryear{{Hartkopf}, {Mason} \& {Rafferty}}{{Hartkopf}
  et~al.}{2008}]{2008AJ....135.1334H}
{Hartkopf} W.~I.,  {Mason} B.~D.,    {Rafferty} T.~J.,  2008, \aj, 135, 1334

\bibitem[\protect\citeauthoryear{{Hartkopf}, {McAlister} \& {Franz}}{{Hartkopf}
  et~al.}{1992}]{1992AJ....104..810H}
{Hartkopf} W.~I.,  {McAlister} H.~A.,    {Franz} O.~G.,  1992, \aj, 104, 810

\bibitem[\protect\citeauthoryear{{Hartkopf}, {McAlister} \& {Mason}}{{Hartkopf}
  et~al.}{2001}]{2001AJ....122.3480H}
{Hartkopf} W.~I.,  {McAlister} H.~A.,    {Mason} B.~D.,  2001, \aj, 122, 3480

\bibitem[\protect\citeauthoryear{{Hartkopf}, {McAlister}, {Mason}, {Barry},
  {Turner} \& {Fu}}{{Hartkopf} et~al.}{1994}]{1994AJ....108.2299H}
{Hartkopf} W.~I.,  {McAlister} H.~A.,  {Mason} B.~D.,  {Barry} D.~J.,  {Turner}
  N.~H.,    {Fu} H.-H.,  1994, \aj, 108, 2299

\bibitem[\protect\citeauthoryear{{Hartkopf}, {McAlister}, {Mason}, {ten
  Brummelaar}, {Roberts} Jr., {Turner} \& {Wilson}}{{Hartkopf}
  et~al.}{1997}]{1997AJ....114.1639H}
{Hartkopf} W.~I.,  {McAlister} H.~A.,  {Mason} B.~D.,  {ten Brummelaar} T.,
  {Roberts} Jr. L.~C.,  {Turner} N.~H.,    {Wilson} J.~W.,  1997, \aj, 114,
  1639

\bibitem[\protect\citeauthoryear{{H{\o}g}, {Fabricius}, {Makarov}, {Urban},
  {Corbin}, {Wycoff}, {Bastian}, {Schwekendiek} \& {Wicenec}}{{H{\o}g}
  et~al.}{2000}]{2000A&A...355L..27H}
{H{\o}g} E.,  {Fabricius} C.,  {Makarov} V.~V.,  {Urban} S.,  {Corbin} T.,
  {Wycoff} G.,  {Bastian} U.,  {Schwekendiek} P.,    {Wicenec} A.,  2000, \aap,
  355, L27

\bibitem[\protect\citeauthoryear{{Horch}, {Bahi}, {Gaulin}, {Howell}, {Sherry},
  {Baena Gall{\'e}} \& {van Altena}}{{Horch}
  et~al.}{2012}]{2012AJ....143...10H}
{Horch} E.~P.,  {Bahi} L.~A.~P.,  {Gaulin} J.~R.,  {Howell} S.~B.,  {Sherry}
  W.~H.,  {Baena Gall{\'e}} R.,    {van Altena} W.~F.,  2012, \aj, 143, 10

\bibitem[\protect\citeauthoryear{{Horch}, {van Altena}, {Howell}, {Sherry} \&
  {Ciardi}}{{Horch} et~al.}{2011}]{2011AJ....141..180H}
{Horch} E.~P.,  {van Altena} W.~F.,  {Howell} S.~B.,  {Sherry} W.~H.,
  {Ciardi} D.~R.,  2011, \aj, 141, 180

\bibitem[\protect\citeauthoryear{{Jeffers}, {van den Bos} \&
  {Greeby}}{{Jeffers} et~al.}{1963}]{1963icvd.book.....J}
{Jeffers} H.~M.,  {van den Bos} W.~H.,    {Greeby} F.~M.,  1963, {Index
  catalogue of visual double stars, 1961.0}.
Vol.~XXI

\bibitem[\protect\citeauthoryear{{Kurucz}}{{Kurucz}}{1994}]{1994KurCD..19.....K}
{Kurucz} R.,  1994, Solar abundance model atmospheres for 0,1,2,4,8
  km/s.~Kurucz CD-ROM No.~19.~ Cambridge, Mass.: Smithsonian Astrophysical
  Observatory, 1994., 19

\bibitem[\protect\citeauthoryear{{Lang}}{{Lang}}{1992}]{1992adps.book.....L}
{Lang} K.~R.,  1992, {Astrophysical Data I. Planets and Stars.}

\bibitem[\protect\citeauthoryear{{Ma{\'{\i}}z-Apell{\'a}niz}}{{Ma{\'{\i}}z-Apell{\'a}niz}}{2004}]{2004PASP..116..859M}
{Ma{\'{\i}}z-Apell{\'a}niz} J.,  2004, \pasp, 116, 859

\bibitem[\protect\citeauthoryear{Ma{\'\i}z-Apell{\'a}niz}{Ma{\'\i}z-Apell{\'a}niz}{2007}]{maiz2007uniform}
Ma{\'\i}z-Apell{\'a}niz J.,  2007, in The Future of Photometric,
  Spectrophotometric and Polarimetric Standardization Vol.~364, A uniform set
  of optical/nir photometric zero points to be used with chorizos.
p.~227

\bibitem[\protect\citeauthoryear{{Malaroda}}{{Malaroda}}{1975}]{1975AJ.....80..637M}
{Malaroda} S.,  1975, \aj, 80, 637

\bibitem[\protect\citeauthoryear{{Martin} \& {Mignard}}{{Martin} \&
  {Mignard}}{1998}]{1998A&A...330..585M}
{Martin} C.,  {Mignard} F.,  1998, \aap, 330, 585

\bibitem[\protect\citeauthoryear{{Mason}, {Hartkopf}, {Gies}, {Henry} \&
  {Helsel}}{{Mason} et~al.}{2009}]{2009AJ....137.3358M}
{Mason} B.~D.,  {Hartkopf} W.~I.,  {Gies} D.~R.,  {Henry} T.~J.,    {Helsel}
  J.~W.,  2009, \aj, 137, 3358

\bibitem[\protect\citeauthoryear{{McAlister}}{{McAlister}}{1978}]{1978ApJ...225..932M}
{McAlister} H.~A.,  1978, \apj, 225, 932

\bibitem[\protect\citeauthoryear{{McAlister} \& {Degioia}}{{McAlister} \&
  {Degioia}}{1979}]{1979ApJ...228..493M}
{McAlister} H.~A.,  {Degioia} K.~A.,  1979, \apj, 228, 493

\bibitem[\protect\citeauthoryear{{McAlister} \& {Fekel}}{{McAlister} \&
  {Fekel}}{1980}]{1980ApJS...43..327M}
{McAlister} H.~A.,  {Fekel} F.~C.,  1980, \apjs, 43, 327

\bibitem[\protect\citeauthoryear{{McAlister} \& {Hartkopf}}{{McAlister} \&
  {Hartkopf}}{1984}]{1984cimb.book.....M}
{McAlister} H.~A.,  {Hartkopf} W.~I.,  1984, {Catalog of interferometric
  measurements of binary stars}

\bibitem[\protect\citeauthoryear{{McAlister}, {Hartkopf}, {Hutter} \&
  {Franz}}{{McAlister} et~al.}{1987}]{1987AJ.....93..688M}
{McAlister} H.~A.,  {Hartkopf} W.~I.,  {Hutter} D.~J.,    {Franz} O.~G.,  1987,
  \aj, 93, 688

\bibitem[\protect\citeauthoryear{{McAlister}, {Hartkopf}, {Sowell},
  {Dombrowski} \& {Franz}}{{McAlister} et~al.}{1989}]{1989AJ.....97..510M}
{McAlister} H.~A.,  {Hartkopf} W.~I.,  {Sowell} J.~R.,  {Dombrowski} E.~G.,
  {Franz} O.~G.,  1989, \aj, 97, 510

\bibitem[\protect\citeauthoryear{{McAlister} \& {Hendry}}{{McAlister} \&
  {Hendry}}{1982a}]{1982ApJS...48..273M}
{McAlister} H.~A.,  {Hendry} E.~M.,  1982a, \apjs, 48, 273

\bibitem[\protect\citeauthoryear{{McAlister} \& {Hendry}}{{McAlister} \&
  {Hendry}}{1982b}]{1982ApJS...49..267M}
{McAlister} H.~A.,  {Hendry} E.~M.,  1982b, \apjs, 49, 267

\bibitem[\protect\citeauthoryear{{McAlister}, {Mason}, {Hartkopf} \&
  {Shara}}{{McAlister} et~al.}{1993}]{1993AJ....106.1639M}
{McAlister} H.~A.,  {Mason} B.~D.,  {Hartkopf} W.~I.,    {Shara} M.~M.,  1993,
  \aj, 106, 1639

\bibitem[\protect\citeauthoryear{{Perryman} \& {et al.}}{{Perryman} \& {et
  al.}}{1997}]{HIPshort}
{Perryman} M.~A.~C.,  {et al.} 1997, \aap, 323, L49

\bibitem[\protect\citeauthoryear{{Salaris} \& {Cassisi}}{{Salaris} \&
  {Cassisi}}{2006}]{2006essp.book.....S}
{Salaris} M.,  {Cassisi} S.,  2006, {Evolution of Stars and Stellar
  Populations}

\bibitem[\protect\citeauthoryear{{Shatskii} \& {Tokovinin}}{{Shatskii} \&
  {Tokovinin}}{1998}]{1998AstL...24..673S}
{Shatskii} N.~I.,  {Tokovinin} A.~A.,  1998, Astronomy Letters, 24, 673

\bibitem[\protect\citeauthoryear{{Tokovinin}}{{Tokovinin}}{2012}]{2012AJ....144...56T}
{Tokovinin} A.,  2012, \aj, 144, 56

\bibitem[\protect\citeauthoryear{{Tokovinin}, {Mason} \&
  {Hartkopf}}{{Tokovinin} et~al.}{2010}]{2010AJ....139..743T}
{Tokovinin} A.,  {Mason} B.~D.,    {Hartkopf} W.~I.,  2010, \aj, 139, 743

\bibitem[\protect\citeauthoryear{{Tokovinin}, {Mason}, {Hartkopf}, {Mendez} \&
  {Horch}}{{Tokovinin} et~al.}{2015}]{2015AJ....150...50T}
{Tokovinin} A.,  {Mason} B.~D.,  {Hartkopf} W.~I.,  {Mendez} R.~A.,    {Horch}
  E.~P.,  2015, \aj, 150, 50

\bibitem[\protect\citeauthoryear{{van Leeuwen}}{{van
  Leeuwen}}{2007}]{2007A&A...474..653V}
{van Leeuwen} F.,  2007, \aap, 474, 653

\bibitem[\protect\citeauthoryear{{Zinnecker} \& {Mathieu}}{{Zinnecker} \&
  {Mathieu}}{2001}]{2001IAUS..200.....Z}
{Zinnecker} H.,  {Mathieu} R.,  eds, 2001, {The Formation of Binary Stars}
  Vol.~200 of IAU Symposium

\end{thebibliography}

%\end{thebibliography}

\end{document}